# Plasma-assisted fabrication of monolayer phosphorene and its Raman characterization

Wanglin Lu[1,§], Haiyan Nan[2,§], Jinhua Hong[1], Yuming Chen[2], Chen Zhu[1], Zheng Liang[3], Xiangyang Ma[1], Zhenhua Ni[2*], Chuanhong Jin[1*], Ze Zhang[1]

[1] State Key Laboratory of Silicon Materials, Key Laboratory of Advanced Materials and Applications for Batteries of Zhejiang Province, Department of Materials Science and Engineering, Zhejiang University, Hangzhou 310027 P. R. China

[2] Department of Physics, Southeast University, Nanjing 211189 P. R. China

[3] Graphene Research and Characterization Center, Taizhou Sunano New Energy Co., Ltd. Taizhou 225300, China

§ These authors contributed equally to this work.

Corresponding authors: zhni@seu.edu.cn; chhjin@zju.edu.cn

## Abstract

There have been continuous efforts to seek for novel functional two-dimensional semiconductors with high performance for future applications in nanoelectronics and optoelectronics. In this work, we introduce a successful experimental approach to fabricate monolayer phosphorene by mechanical cleavage and the following $Ar^+$ plasma thinning process. The thickness of phosphorene is unambiguously determined by optical contrast combined with atomic force microscope (AFM). Raman spectroscopy is used to characterize the pristine and plasma-treated samples. The Raman frequency of $A^2_g$ mode stiffens, and the intensity ratio of $A^2_g$ to $A^1_g$ modes shows monotonic discrete increase with the decrease of phosphorene thickness down to monolayer. All those phenomena can be used to identify the thickness of this novel two-dimensional semiconductor efficiently. This work for monolayer phosphorene fabrication and thickness determination will facilitates the research of phosphorene.

**Introduction**

Since the first discovery of graphene, there have been a lot of advances achieved on the fabrication of novel two-dimensional crystals down to unit-cell thickness via mechanical cleavage, which has opened up new possibilities for future nanoelectronics and optoelectronics[1-3]. Among them, the exfoliated graphene and $MoS_2$ are two successful representatives[4-6]. Graphene exhibits ultra-high carrier mobility owing to its massless Dirac feature[7], while the lack of band gap severely limits its application in logic semiconductor devices like field-effect transistors. $MoS_2$, a member of transition metal dichalcogenides family, is a direct-band gap semiconductor in the form of monolayer[8]. Transistors made of $MoS_2$ have shown device performance with a mobility of 200 $cm^2/V·s$[6, 9, 10], which might be limited by the presence of structural defects [11]. To seek for novel functional two-dimensional semiconductor is still on demanding.

Black phosphorus, the most stable form among the various allotropic modifications of phosphorus, is a layered solid stacked with atomic layers via weak van der Waals interactions (Figure 1a) similar to graphite[12]. Early studies show that bulk black phosphorus is a semiconductor with unique electronic and optical properties [13]. Very recently, Zhang and co-workers report that few-layer black phosphorus based field-effect transistors exhibit excellent electronic properties with a high on-off ratio of $10^6$ and a carrier mobility of ca. 1000 $cm^2/V·s$[14]. Since thickness is one of the critical parameters that defines the electronic, optical, and thermal properties of two-dimensional crystals, it is natural to ask whether we can achieve monolayer phosphorene.

In this work, we present the first successful experimental approach to fabricate monolayer phosphorene by mechanical cleavage and plasma thinning. The thickness is unambiguously determined by optical contrast combined with AFM. Raman spectroscopy is used to characterize the pristine and plasma-treated samples. By comparing the Raman spectra collected from mono- to few-layer phosphorene, we find that the in-plane $A^2_g$ mode exhibits thickness-sensitive frequency shift, while the out-of-plane $A^1_g$ and in-plane $B_{2g}$ modes are almost unaffected by thickness. The optical contrast and the integrated Raman intensity ratio of $A^2_g$ to $A^1_g$ increase monotonically with the increase of layer numbers, and therefore can be used as reliable fingerprints for determining the thickness of phosphorene.

**Results and Discussion**

Thin phosphorene films are firstly exfoliated onto $SiO_2$/Si substrate by mechanical exfoliation and identified by optical microscopy. The sample thickness is further determined by AFM combined with contrast spectra (discussed later). With hundreds of few-layer sheets checked, the thinnest forms we can get are bilayer flakes. The reason why monolayer phosphorene cannot be obtained easily via mechanical exfoliation is still unclear here. It might be due to the strong inter-layer coupling so that multilayer flakes are easier accessed. It is therefore critical to find a simple and efficient method to fabricate monolayer phosphorene. Recently, it has been demonstrated that layer-by-layer thinning of multilayers of $MoS_2$ under plasma or laser irradiation is an efficient approach to get monolayer $MoS_2$[15, 16]. Here, we demonstrate plasma thinning also works as an effective way to achieve monolayer phosphorene. Figure 1b and 1c show optical images of thin sheets of phosphorene deposited on $SiO_2$/Si substrate before and after $Ar^+$ plasma treatment, respectively. As can be seen, the optical contrast of the sample decreases obviously and some of the regions become single layer after plasma thinning (as identified in Figure 2 later). The optical reflection spectrum image[17] (Figure 1d) of the same sample after plasma treatment is very uniform for different regions with variant thickness, which shows that plasma thinning is a highly controllable technique to get homogeneous monolayer and few-layers phosphorene. Figure 1e is a typical TEM image showing the morphology of few-layer phosphorene after plasma thinning, and the inset selected-area electron diffraction pattern indicates they are highly-crystalline.

Figure 2a shows the contrast spectra for 1-5 layers of phosphorene. The optical contrast is defined as $(R_o-R)/R_o$, where $R_o$ and R are the reflection spectrum from the substrate and samples, respectively. Each spectrum has a peak centered at 550 nm. The highest contrast values for 1-5 layers are 0.16, 0.32 0.41, 0.47 and 0.50 respectively (Figure 2b), which is convenient for thickness determination in the further study of phosphorene. Two phosphorene flakes with optical contrast of ~ 0.16 and ~ 0.5 are imaged by AFM (insets in Figure 2b). The measured heights are 0.85 nm and 2.8 nm respectively which are well consistent with the heights of mono-layer and five layers of phosphorene. It also agrees well with the thickness identification by optical contrast. The optical contrast value of phosphorene is larger compared to that of graphene, which may imply that the optical absorption of phosphorene is stronger than that of graphene[18].

Figure 3a shows the Raman spectra of the pristine exfoliated samples with thickness of 2-5 layers. We observe one out-of-plane modes ($A^1_g$) and two in-plane modes ($A^2_g$ and $B_{2g}$) (Figure 3d)[19]. We also measure the Raman spectra of plasma-treated sheets with thickness of 1-5 layers (Figure 3b), and compared with those of pristine sheets. The Raman characteristics (peak frequencies and intensity ratios) of pristine and plasma-treated samples are almost identical as shown in Figure 3c and Figure 4c, suggesting that plasma treatment does not introduce noticeable structural defects on thin phosphorene, similar to the results reported for plasma-thinned $MoS_2$[15]. This again validates the advantage for plasma thinning to scale up the generation of phosphorene with well-controlled number of layers. For both pristine and plasma- treated films, the $A^2_g$ mode shifts from 467.7 cm$^{-1}$ (bulk) to 470 cm$^{-1}$ (bilayer), while $A^1_g$ and $B_{2g}$ modes remain unchanged at ~440 cm$^{-1}$ and ~363 cm$^{-1}$ with the decrease of thickness. Most importantly, the frequency of $A^2_g$ mode of monolayer phosphorene further shifts to

471.3 cm$^{-1}$, while B$_{2g}$ still remains at ~440 cm$^{-1}$, which are consistent with the trend of peak frequency shift with thickness for pristine samples. The frequency difference between A$^2_g$ and B$_{2g}$ modes changes from 27.7 cm$^{-1}$ of bulk to 31 cm$^{-1}$ of monolayer (Figure 3c), which could be an effective parameter to identify the thickness of phosphorene.

The in-plane mode A$^2_g$ vibration stiffens with the decrease of thickness, while B$_{2g}$ and A$^1_g$ modes almost keep unchanged. All those vibrations go against the prediction for the coupled harmonic oscillator, within which those modes are expected to stiffen with the increase of thickness because of the additional interlayer van der Waals interaction. In the case of MoS$_2$, such an anomalous vibration of E$^1_{2g}$ mode is attributed to the dielectric screening mainly due to the presence of molybdenum atoms[20, 21]. Note that only phosphorus atoms are involved in vibration, dielectric screening should be negligible while stacking induced structure change may dominate. Black phosphorus is an unique anisotropic material with the lattice parameters in different directions exhibiting different sensitiveness to external impacts[22, 23]. Theoretical calculations show that lattice parameter along z direction (as shown in Figure 3d) changes significantly from bulk form to monolayer phosphorene while other two directions almost remain unchanged [24]. Those stacking-induced anisotropic structure changes may help explain these anomalous behaviors of vibrations.

Figure 3b shows that the Raman peak intensities of bulk phosphorus are smaller than those of multilayer phosphorene, similar to MoS$_2$ and graphene, which is due to the effects of optical interference[25, 26]. Figure 4a and 4b present the Raman mappings of A$^2_g$ and A$^1_g$ peak intensities, respectively, with higher intensities for thicker regions. The integrated intensity ratio of A$^2_g$ and A$^1_g$ presents monotonically discrete increase from bulk to monolayer phosphorene (Figure 4c). The ratio changes mildly from bulk (1.9) to triple-layer (3.5) while drastic jump from triple- to bi-layer (7.8±0.6) and mono-layer (10.5±1). According to theoretical predictions[24, 27], the anisotropic structure changes of phosphorene with different thickness may induce anisotropic variation of electronic structure and optical properties (e.g. Raman scattering), so the changes of intensity ratio as a function of thickness is possible. Figure 4d presents two typical Raman spectra including the first order (Si) peak at ~520 cm$^{-1}$. We find that the intensity ratios of $A^2_g$ and first order (Si) peak are monotonic discrete increase from monolayer to penta-layers due to the enhanced optical absorption of thicker flakes. The distinctive ratios discussed above combined with frequency features can be used as effective and reliable thickness indicators.

Regarding the possibility of transformation between white, red and black phosphorus, we preserve the thin films including monolayer at dry atmosphere several months and annealed in 200℃ in good vacuum condition of $2\times10^{-4}$ Pa. We find that few-layer phosphorene still remain high crystalline after annealing. Those results light up the promising application of phosphorene in electronic and optical devices.

**Conclusion**

In summary, we propose the first successful experimental approach to achieve stable monolayer phosphorene through a combination of mechanical cleavage and Ar$^+$ plasma treatment. The thickness of thin flakes (up to five here) can be rigidly identified by optical contrast and AFM. We have characterized pristine and plasma-treated thin films by Raman spectroscope. The frequency of $A^2_g$ mode is found to stiffen while $B_{2g}$ and $A^1_g$ modes remain unchanged with the decrease of thickness, which is probably due to the influence of stacking-induced anisotropic structure change. We also observe dramatic anisotropic intensity changes of inter- and intra-plane modes. Those features enable rapid and unambiguous identification of thin phosphorene flakes by Raman spectroscopy. The results presented here will play an essential role for fabricating high-performance phosphorene based nanoelectronic devices.

**Methods**

High-quality black phosphorus crystals were purchased from Smart-Element. Few-layer phosphorene sample was obtained by mechanical cleavage onto Si substrate with a 300 nm $SiO_2$ capping layer. $Ar^+$ plasma (commercial 13.56 MHz RF source) with a power of 30 W and a pressure of 30 Pa was used to thin down the thin phosphorene sheets with 20 s at room temperature. Tapping mode AFM was done on a Bruker Dimension Edge Microscope.

The reflection/contrast and Raman spectra were carried out with a WITec alpha 300R confocal Raman system. In the reflection/contrast experiments, the incident light was emitted from a normal white light source and collected using backscattering configuration. The excitation source for Raman measurement is 532 nm laser (2.33 eV) with a laser power below 0.15mW on the samples to avoid laser induced heating. A 100× objective lens with a NA=0.90 was used both in the Raman and reflection experiments, and the spot size of 532 nm laser and white light were estimated to be 300 nm and 1 um, respectively. For the reflection and Raman images, the sample was placed on an x-y piezostage and scanned under the illumination of laser and white light. The Raman and reflection spectra from every spot of the sample were recorded and analyzed by WITec Project software

The TEM samples were prepared by transferring the thin phosphorene onto the conventional lacey-carbon TEM supports after etching away the underneath $SiO_2$ layer with concentrated potassium hydroxide solution. TEM characterizations were carried out with a FEI Tecnai $G^2$-F20 microscope operated at 200 kV.

## Acknowledgements

The authors would like to thank Prof. Wei Ji from Renmin University for his kindness on sharing with us the unpublished results on electronic structure calculations of phosphorene, Prof. Pingheng Tan for his guide on early Raman characterizations and Dr. Shuo Ding for her assistance to obtain the optical images used in TOC. This work is financially supported by the National Science Foundation of

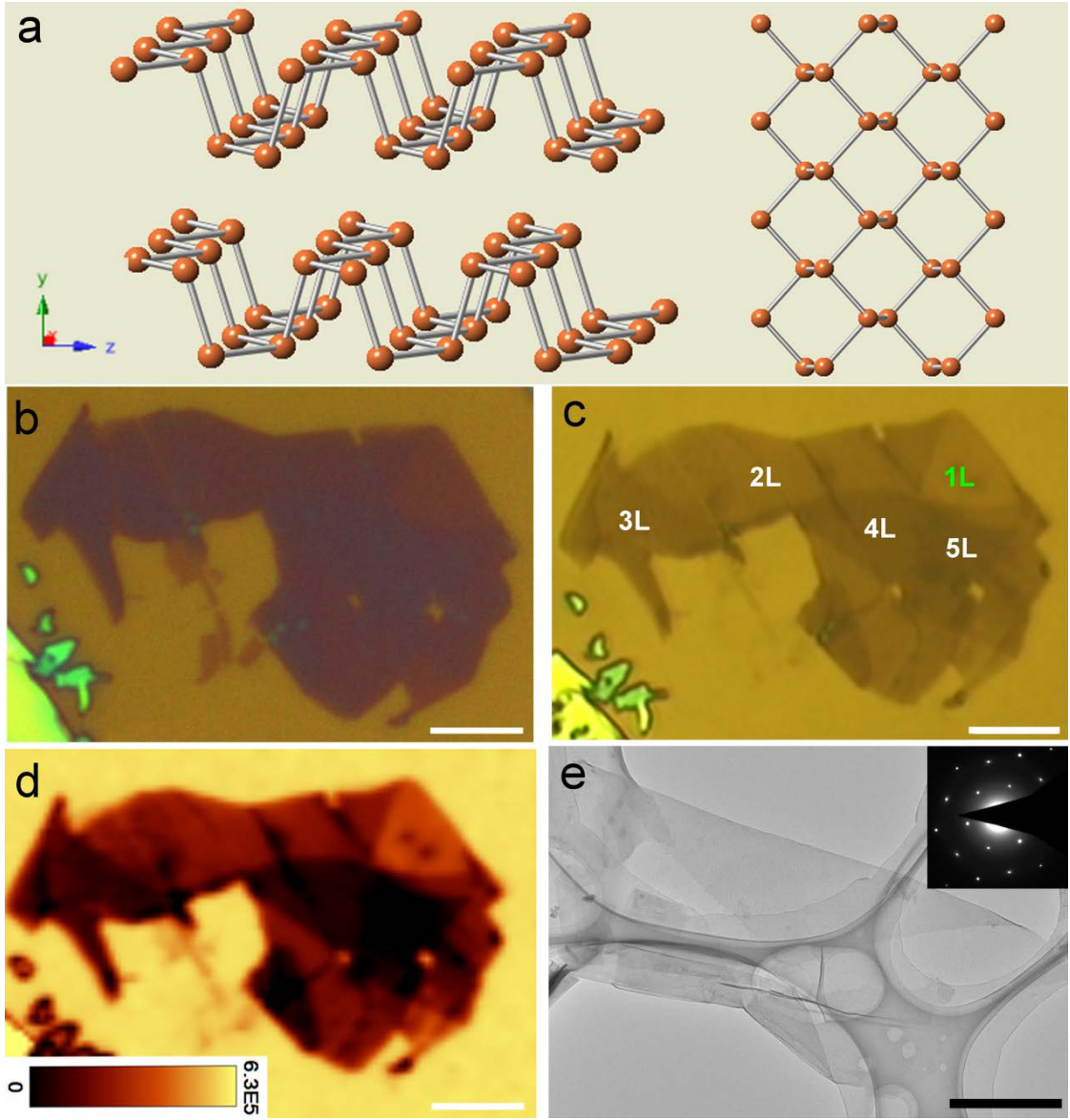

**Figure 1.** (a) Perspective side view (left) and top view (right) of black phosphorus. (b) Optical image of multilayered pristine phosphorene. (c) Same as in (a) after Ar$^+$ plasma thinning. (d) Reflection image of plasma treated flake in (b). Scale bars in b, c and d are 5 um. (e) TEM image of few-layer phosphorene with an inset showing the SAED pattern (scale bar: 500nm).

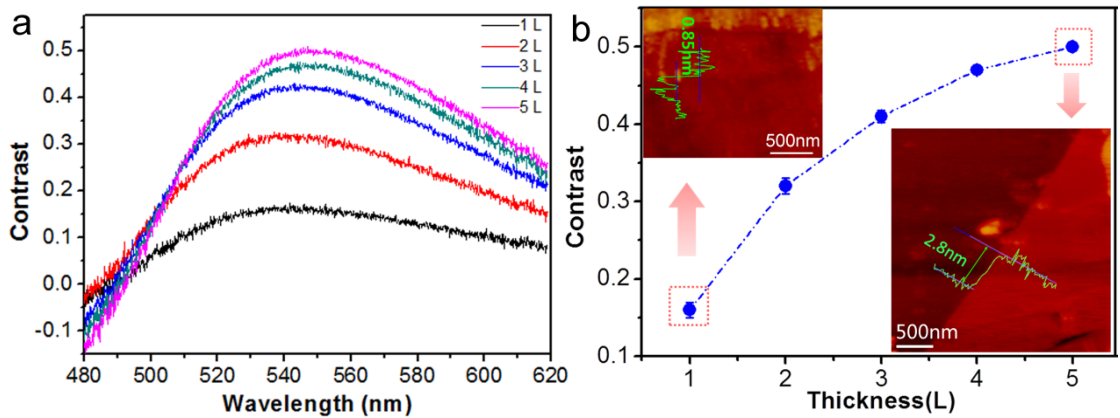

**Figure 2.** (a) The contrast spectra of phosphorene with different layers. (b) Contrast values of 1-5 layer phosphorene (inset: The corresponding AFM images from two phosphorene flakes with different optical contrast).

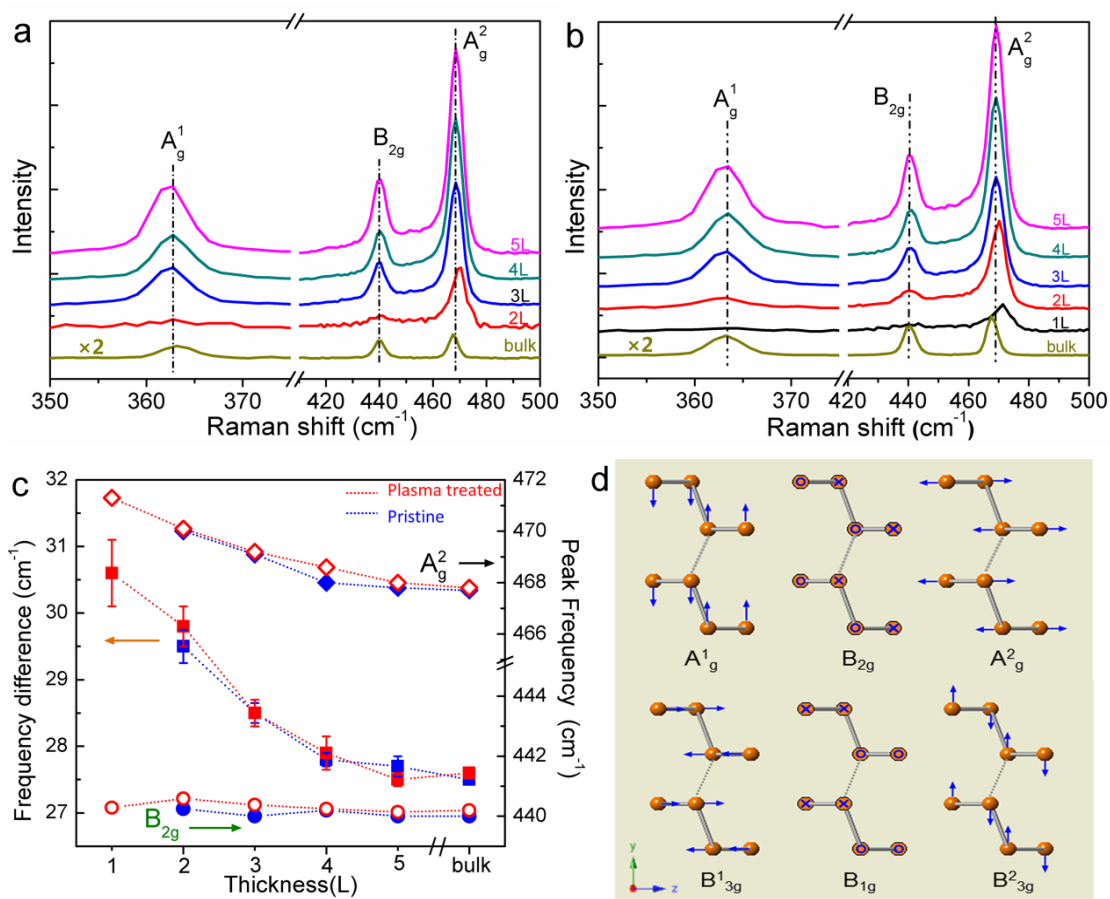

**Figure 3.** (a) Raman spectra of pristine phosphorene with different number of layers. (b) Raman spectra of plasma-treated thin and bulk phosphorene. (c) Peak frequency of $A^2_g$ (up) and $B_{2g}$ (bottom) Raman modes (right vertical axis) and the difference between them (left vertical axis) with thickness changing. (d) Atomic displacement of Raman active modes along x direction.

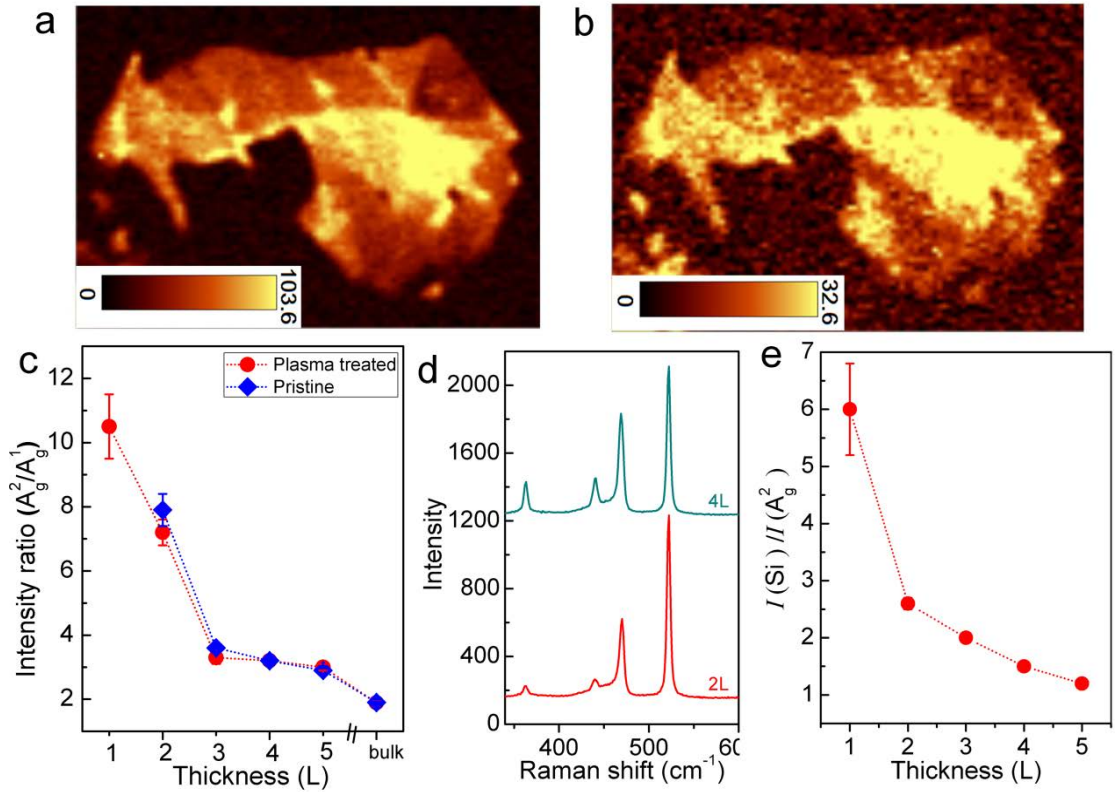

**Figure 4.** Raman mapping images of $A^2_g$ (a) and $A^1_g$ (b) peak intensities. (c) The intensity ratio of $A^2_g$ and $A^1_g$. (d) Raman spectra of bi-layer and tetra-layer phosphorene including the first order (Si) peak of the silicon substrate. (e) The intensity ratio of $A^2_g$ and first order (Si) peak.